\documentclass[12pt]{article}

\usepackage{cite}
\usepackage{epsfig}

\def\beq{\begin{equation}}
\def\eeq{\end{equation}}
\def\beqar{\begin{eqnarray}}
\def\eeqar{\end{eqnarray}}
\def\barr#1{\begin{array}{#1}}
\def\earr{\end{array}}
\def\bfi{\begin{figure}}
\def\efi{\end{figure}}
\def\btab{\begin{table}}
\def\etab{\end{table}}
\def\bce{\begin{center}}
\def\ece{\end{center}}

\def\text{\textstyle}

\def\al{\alpha}

\def\ga{\gamma}
\def\de{\delta}

\def\Ga{\Gamma}
\def\De{\Delta}

\def\refeq#1{\mbox{eq.~(\ref{#1})}}
\def\refeqs#1{\mbox{eqs.~(\ref{#1})}}
\def\reffi#1{\mbox{Fig.~\ref{#1}}}

\def\refFi#1{\mbox{Figure~\ref{#1}}}

\def\refta#1{\mbox{Table~\ref{#1}}}

\def\citere#1{\mbox{Ref.~\cite{#1}}}
\def\citeres#1{\mbox{Refs.~\cite{#1}}}

\newcommand{\GeV}{\unskip\,\mathrm{GeV}}
\newcommand{\MeV}{\unskip\,\mathrm{MeV}}

\def\mathswitchr#1{\relax\ifmmode{\mathrm{#1}}\else$\mathrm{#1}$\fi}

\newcommand{\PW}{\mathswitchr W}
\newcommand{\PZ}{\mathswitchr Z}

\newcommand{\PH}{\mathswitchr H}

\newcommand{\Pt}{\mathswitchr t}

\def\mathswitch#1{\relax\ifmmode#1\else$#1$\fi}

\newcommand{\MW}{\mathswitch {M_\PW}}

\newcommand{\MZ}{\mathswitch {M_\PZ}}
\newcommand{\MH}{\mathswitch {M_\PH}}

\newcommand{\Mt}{\mathswitch {m_\Pt}}
\newcommand{\scrs}{{}}
\newcommand{\sw}{\mathswitch {s_{\scrs\PW}}}
\newcommand{\swtwo}{\mathswitch {s_{{\scrs\PW}, (2)}}}

\newcommand{\cw}{\mathswitch {c_{\scrs\PW}}}
\newcommand{\sweff}{\sin^2 \theta_{\mathrm{eff}}}

\newcommand{\MWsub}{M_{\PW, \mathrm{subtr}}}

\newcommand{\GF}{\mathswitch {G_\mu}}

\newcommand{\mt}{\Mt}

\newcommand{\tsf}{\theta\kern-.20em_{\tilde{f}}}
\newcommand{\tsfp}{\theta\kern-.20em_{\tilde{f}\prime}}
\newcommand{\tsq}{\theta\kern-.15em_{\tilde{q}}}

\newcommand{\lsim}
{\;\raisebox{-.3em}{$\stackrel{\displaystyle <}{\sim}$}\;}

\newcommand{\alps}{\alpha_{\mathrm s}}

\newcommand{\fea}{{\em FeynArts}}

\newcommand{\two}{{\em TwoCalc}}

\def\rT{{\mathrm{T}}}

\newcommand{\msbar}{$\overline{\rm{MS}}$}

\newcommand{\xiaz}{\xi^{\gamma \PZ}}
\newcommand{\xiza}{\xi^{\PZ\gamma}}

\newcommand{\VL}{\left( \begin{array}{c}}
\newcommand{\VR}{\end{array} \right)}
\newcommand{\ML}{\left( \begin{array}{cc}}
\newcommand{\MLd}{\left( \begin{array}{ccc}}
\newcommand{\MLv}{\left( \begin{array}{cccc}}
\newcommand{\MR}{\end{array} \right)}

\newcommand{\tev}{\,\, \mathrm{TeV}}
\newcommand{\gev}{\,\, \mathrm{GeV}}

\newcommand{\BC}{\begin{center}}
\newcommand{\EC}{\end{center}}
\newcommand{\BE}{\begin{equation}}
\newcommand{\EE}{\end{equation}}
\newcommand{\BEA}{\begin{eqnarray}}
\newcommand{\BEAnn}{\begin{eqnarray*}}
\newcommand{\EEA}{\end{eqnarray}}
\newcommand{\EEAnn}{\end{eqnarray*}}
\newcommand{\non}{\nonumber}
\newcommand{\id}{{\rm 1\kern-.12em
\rule{0.3pt}{1.5ex}\raisebox{0.0ex}{\rule{0.1em}{0.3pt}}}}

\def\draftdate{\relax}
\def\mda{\relax}
\def\mua{\relax}
\def\mla{\relax}
\def\draft{
\def\thtystars{******************************}
\def\sixtystars{\thtystars\thtystars}
\typeout{}
\typeout{\sixtystars**}
\typeout{* Draft mode!
         For final version remove \protect\draft\space in source file
*}
\typeout{\sixtystars**}
\typeout{}
\def\draftdate{\today}
\def\mua{\marginpar[\boldmath\hfil$\uparrow$]%
                   {\boldmath$\uparrow$\hfil}%
                    \typeout{marginpar: $\uparrow$}\ignorespaces}
\def\mda{\marginpar[\boldmath\hfil$\downarrow$]%
                   {\boldmath$\downarrow$\hfil}%
                    \typeout{marginpar: $\downarrow$}\ignorespaces}
\def\mla{\marginpar[\boldmath\hfil$\rightarrow$]%
                   {\boldmath$\leftarrow $\hfil}%
                    \typeout{marginpar:
$\leftrightarrow$}\ignorespaces}
\def\Mua{\marginpar[\boldmath\hfil$\Uparrow$]%
                   {\boldmath$\Uparrow$\hfil}%
                    \typeout{marginpar: $\Uparrow$}\ignorespaces}
\def\Mda{\marginpar[\boldmath\hfil$\Downarrow$]%
                   {\boldmath$\Downarrow$\hfil}%
                    \typeout{marginpar: $\Downarrow$}\ignorespaces}
\def\Mla{\marginpar[\boldmath\hfil$\Rightarrow$]%
                   {\boldmath$\Leftarrow $\hfil}%
                    \typeout{marginpar:
$\Leftrightarrow$}\ignorespaces}
\overfullrule 5pt
\oddsidemargin -15mm
\marginparwidth 29mm
}

\begin{document}
\thispagestyle{empty}

\def\thefootnote{\fnsymbol{footnote}}

\begin{flushright}
CERN--TH/2000--194\\
DESY 00--097\\
KA-TP--13--2000\\
hep-ph/0007091 \\
\end{flushright}

\vspace{1cm}

\begin{center}

{\Large\sc {\bf Complete fermionic two-loop results for the\\[.5em]
 $\MW$--$\MZ$ interdependence }}\\[3.5em]
{\large
{\sc
A.~Freitas$^{1}$, W.~Hollik$^{2}$, W.~Walter$^{2}$,
and G.~Weiglein$^{3}$%
\footnote{email: Georg.Weiglein@cern.ch}
}
}

\vspace*{1cm}

{\sl
$^1$ DESY Theorie, Notkestr. 85, D--22603 Hamburg, Germany

\vspace*{0.4cm}

$^2$ Institut f\"ur Theoretische Physik, Universit\"at Karlsruhe, \\
D--76128 Karlsruhe, Germany

\vspace*{0.4cm}

$^3$ CERN, TH Division, CH--1211 Geneva 23, Switzerland
}

\end{center}

\vspace*{2.5cm}

\begin{abstract}
The complete fermionic two-loop contributions to the prediction for the 
W-boson mass from muon decay in the electroweak Standard Model are evaluated 
exactly, i.e.\ no expansion in the top-quark and the Higgs-boson mass is
made. The result for the W-boson mass is compared with the previous
result of an expansion up to next-to-leading order in the top-quark
mass. The predictions are found to agree with each other within 
about $5$~MeV.
A simple parametrization of the new result is presented, approximating
the full result to better than 0.4~MeV for $\MH \leq 1$~TeV.
\end{abstract}

\def\thefootnote{\arabic{footnote}}
\setcounter{page}{0}
\setcounter{footnote}{0}

\newpage


The prediction of the W-boson mass, $\MW$, in terms of the Z-boson
mass, $\MZ$, the Fermi constant, $\GF$, and the fine structure constant,
$\al$, is one of the most important quantities for testing the
electroweak Standard Model (SM) and its extensions with high precision.
This relation is derived from muon decay, as the Fermi constant is
defined in terms of the muon lifetime, $\tau_{\mu}$, according to
\beq
\tau_{\mu}^{-1} = \frac{\GF^2 \, m_\mu^5}{192 \pi^3} \;
F\left(\frac{m_{\mathrm{e}}^2}{m_\mu^2}\right)
\left(1 + \frac{3}{5} \frac{m_\mu^2}{\MW^2} \right) 
\left(1 + \Delta q \right) ,
\label{eq:fermi}
\eeq
with $F(x) = 1 - 8 x - 12 x^2 \ln x + 8 x^3 - x^4$. By convention, the
QED corrections within the Fermi Model, $\Delta q$, are included in this
defining equation for $\GF$. The one-loop result for 
$\Delta q$~\cite{delqol}, 
which has already been known for several
decades, has recently been supplemented by the two-loop
correction~\cite{delqtl}. The tree-level W~propagator effects giving
rise to the (numerically insignificant) term $3 m_\mu^2/(5 \MW^2)$ in 
\refeq{eq:fermi} are conventionally also included in the definition of
$\GF$, although they do not belong to the Fermi Model prediction.

Comparing the prediction for the muon lifetime within the SM with 
\refeq{eq:fermi} yields the relation
\beq
\MW^2 \left(1 - \frac{\MW^2}{\MZ^2}\right) = 
\frac{\pi \al}{\sqrt{2} \GF} \left(1 + \De r\right),
\label{eq:delr}
\eeq
where the radiative corrections are summarized 
in the quantity $\De r$~\cite{sirlin}.
This relation can be used for deriving the prediction of $\MW$ within 
the SM or extensions of it, to be confronted with the experimental 
result for $\MW$. At present, the W-boson mass is measured with an
accuracy of $5 \times 10^{-4}$, 
$\MW^{\mathrm{exp}} = 80.419 \pm 0.038$~GeV~\cite{mori00}.
The experimental precision on $\MW$
will be further improved with the data taken at LEP2 in its
final year of running, and at the upgraded Tevatron and the LHC, where 
an error of $\de\MW = 15$~MeV can be expected~\cite{lhctdr}. At a 
high-luminosity linear collider running in a low-energy mode at the 
$\PW^+\PW^-$ threshold, a reduction 
of the experimental error down to 
$\de\MW = 6$~MeV can be envisaged~\cite{gigazMW}. This offers the
prospect for highly sensitive tests of the electroweak
theory~\cite{gigaztests}, provided that the accuracy of the theoretical
prediction matches the experimental precision.

The one-loop result for $\De r$ within the SM~\cite{sirlin} 
can be decomposed as (with $\sw^2 = 1 - \MW^2/\MZ^2$)
\beq
\De r^{(\al)} = \De \al - \frac{\cw^2}{\sw^2} \De\rho + 
\De r_{\mathrm{rem}}(\MH),
\label{eq:delrol}
\eeq
where the leading fermion-loop contributions $\De \al$ and $\De\rho$,
arising from the charge and mixing-angle renormalization,
are separated out, while the remainder part $\De r_{\mathrm{rem}}$ contains 
in particular the dependence on the Higgs-boson mass, $\MH$. 
The QED-induced shift in the fine structure constant, $\De \al$, contains 
large logarithms of light-fermion masses. 
The leading contribution to the $\rho$~parameter from
the top/bottom weak isospin doublet, $\De\rho$, gives rise to a term
with a quadratic dependence on the top-quark mass, $\mt$~\cite{velt}.

Beyond the one-loop order, resummations of the leading one-loop
contributions $\De\al$ and $\De\rho$ are known~\cite{resum}.
They correctly take into account the terms of the form 
$(\De\al)^2$, $(\De\rho)^2$, $(\De\al\De\rho)$, and
$(\De\al\De r_{\mathrm{rem}})$ at the two-loop level and the leading
powers in $\De\al$ to all orders. 

While the QCD corrections to $\De r$ are known at ${\cal O}(\al
\alps)$~\cite{qcd2} and ${\cal O}(\al \alps^2)$~\cite{qcd3}, only
partial results are available up to now for the electroweak two-loop
contributions. They have been obtained using expansions for
asymptotically large values of $\mt$~\cite{ewmt4,ewmt2} and 
$\MH$~\cite{ewmh2}. The terms derived by expanding in the top-quark
mass of ${\cal O}(\GF^2 \Mt^4)$~\cite{ewmt4} and 
${\cal O}(\GF^2 \Mt^2 \MZ^2)$~\cite{ewmt2} were found to be numerically
sizeable. The ${\cal O}(\GF^2 \Mt^2 \MZ^2)$ term, involving three different 
mass scales, has been obtained by two separate expansions in the regions
$\MW, \MZ, \MH \ll \mt$ and $\MW, \MZ \ll \mt, \MH$ and by an interpolation
between the two expansions. This formally next-to-leading order term
turned out to be of a magnitude similar to that of
the formally leading term of ${\cal O}(\GF^2 \Mt^4)$, entering with
the same sign. Its inclusion (both for $\MW$ and the effective mixing
angle $\sweff$) had important consequences on the indirect constraints
on the Higgs-boson mass derived from the SM fit to the precision data.

Consequently, a more complete calculation of electroweak two-loop
effects appears desirable, where no expansion in $\mt$ or $\MH$ is
made. As a first step in this direction, exact results have been
obtained for the Higgs-mass dependence (e.g.\ the quantity 
$\MWsub(\MH) \equiv \MW(\MH) - \MW(\MH = 65 \gev))$ of the fermionic two-loop 
corrections to the precision observables~\cite{ewmhdep}. They have been
compared with the results of expanding up to ${\cal O}(\GF^2 \Mt^2
\MZ^2)$~\cite{ewmt2}, specifically analysing the effects of the $\mt$
expansion, and good agreement has been found~\cite{gsw}.

Beyond the two-loop order, complete results for the pure fermion-loop 
corrections (i.e.\ contributions containing $n$ fermion loops at 
$n$-loop order) have recently been obtained up to four-loop order~\cite{floops}.
These results contain in particular the contributions of the leading
powers in $\De\al$ as well as the ones in $\De\rho$ and the mixed
terms.

\begin{figure}[ht]
\vspace{1em}
\begin{center}
\psfig{figure=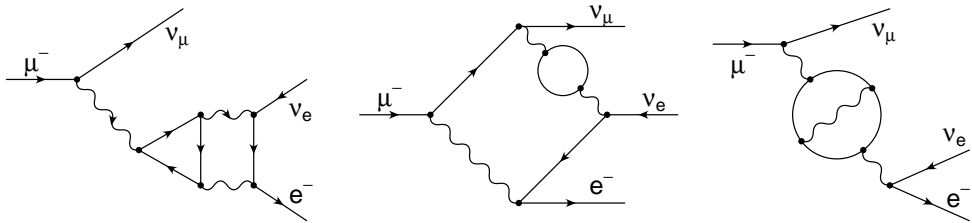,width=13cm}
\vspace{-1em}
\end{center}
\caption[]{{\small
Examples for types of fermionic two-loop diagrams contributing to muon
decay.
\label{fig:diags}
}}
\end{figure}

In the present paper, all fermionic two-loop corrections to $\De r$ are
calculated exactly, i.e.\ without an expansion in the top-quark or the
Higgs-boson mass. These are all two-loop diagrams contributing to 
the muon decay amplitude and
containing at least one closed fermion loop (except the pure QED
corrections already contained in the Fermi model result, see
\refeq{eq:fermi}). 
{\refFi{fig:diags} displays some typical examples}.
The considered class of diagrams includes the
potentially large corrections both from the top/bottom doublet and 
from contributions proportional to $N_{\mathrm{lf}}$ and $N_{\mathrm{lf}}^2$, 
where $N_{\mathrm{lf}}$ is the number of light fermions (a partial result for 
the light-fermion contributions has been obtained in \citere{stuartlf}). The 
results presented here improve on the previous results of an expansion in $\mt$
up to next-to-leading order~\cite{ewmt2} in containing the full
dependence on $\mt$ as well as the complete light-fermion contributions
at the two-loop order, while in \citere{ewmt2} higher-order corrections from
light fermions have only been taken into account via a resummation of
the one-loop light-fermion contribution.

\bigskip
In the following, we briefly outline the main features of the calculation.  
After extracting the IR-divergent
QED corrections that are already contained in the Fermi model QED
factor (a detailed description of how this is done will be given in a
forthcoming paper), the generic diagrams contributing to muon decay can be 
reduced to vacuum-type diagrams, 
since the masses of the external particles  and
the momentum transfer are negligible.
The on-shell renormalization of the gauge-boson masses, on the other
hand, requires the evaluation of two-loop two-point functions with non-zero 
external momentum, which is more involved from a technical point of view
regarding the tensor structure and the evaluation of the scalar
integrals. It should be noted that this complication cannot be avoided by 
performing the calculation within another renormalization scheme (the \msbar\
scheme, for instance), since ultimately one is interested in the relation
between the physical parameters $\MW$, $\MZ$, $\al$, $\GF$, rather than
between their \msbar\ counterparts. For this reason we have decided to
use the on-shell renormalization scheme everywhere in our 
calculation, i.e.\ we use physical parameters throughout (alternatively 
one could of course do the calculation in a 
different renormalization scheme, with formal parameters, 
and perform the transition to the physical parameters in a 
second step). If not otherwise stated, we use the conventions of
\citere{Dehabil}.

In our calculation we have made use of some computer-algebra tools. The
package \fea~\cite{fea} was applied to generate the Feynman
amplitudes and counterterm contributions. The program \two~\cite{two}
was applied for the algebraic evaluation of these amplitudes, which 
were reduced, by means of two-loop tensor-integral decompositions, to 
a set of standard scalar integrals. The calculation was carried out in a
general $R_{\xi}$~gauge, which allowed us to test the gauge-parameter
independence at the algebraic level as a highly non-trivial check. For
the evaluation of the scalar one-loop integrals and the two-loop 
\mbox{vacuum} integrals we have used analytical results as given in
\citere{davtausk}, while the
two-loop two-point integrals with non-vanishing external momentum have 
been evaluated numerically using one-dimensional integral
representations with elementary functions~\cite{intnum}. These allow a
very fast calculation of the integrals for general mass configurations.

\begin{figure}[ht]
\vspace{1em}
\begin{center}
\psfig{figure=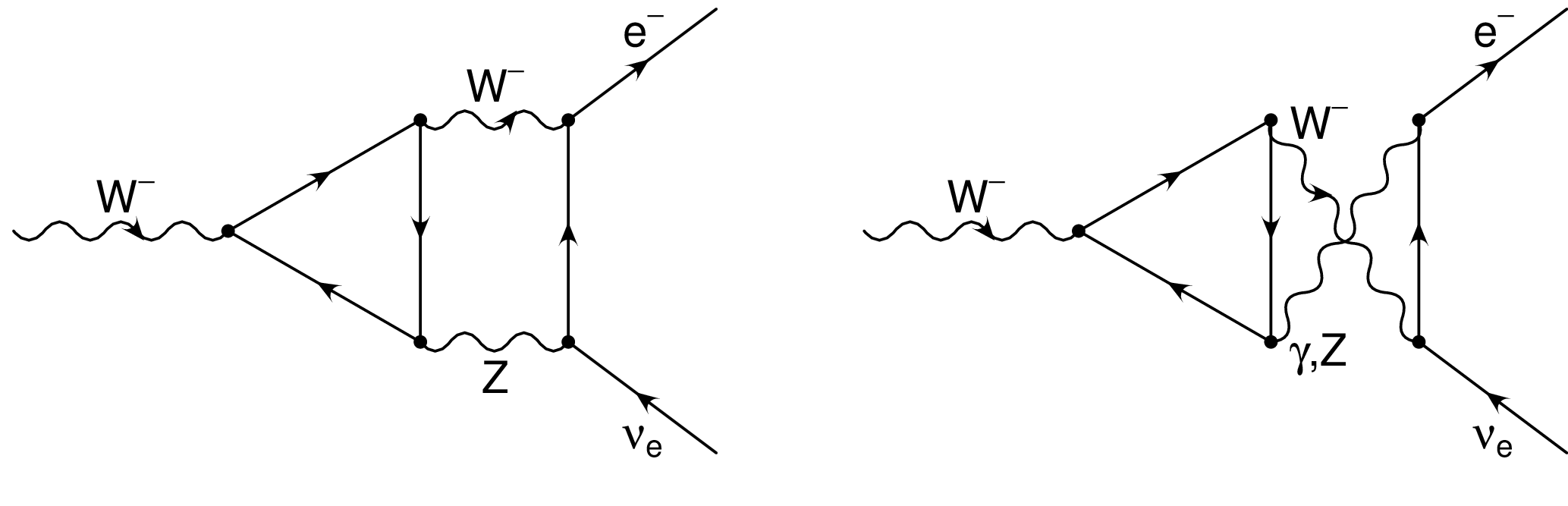,width=10cm}
\vspace{-1em}
\end{center}
\caption[]{{\small
Two-loop vertex diagrams containing a triangle subgraph, which require a
careful treatment of $\ga_5$ in $D$ dimensions.
\label{fig:graphsga5}
}}
\end{figure}

Since we are using Dimensional Regularization~\cite{dreg,ga5HV} in our
calculation, a careful treatment of the Dirac algebra in $D$ dimensions
involving $\ga_5$ is necessary. While a naively anticommuting $\ga_5$
can safely be applied for all two-loop two-point contributions (for a
discussion, see e.g.\ the first paper of~\citere{ewmt4}) and most of the
two-loop vertex- and box-type diagrams, this is not the case for the two-loop
vertex diagrams containing a triangle subgraph, shown in
\reffi{fig:graphsga5}. For these graphs, a naively
anticommuting $\ga_5$, although respecting the Ward identities, would
lead to an incorrect result. This is due to an inconsistent evaluation of 
the trace of $\ga_5$ together with four Dirac matrices, which in four 
dimensions is given by $\mathrm{Tr}\left\{
\gamma_5 \gamma^\mu\gamma^\nu\gamma^\rho\gamma^\sigma \right\} = 4i
\epsilon^{\mu\nu\rho\sigma}$, while applying the naively 
anticommuting $\ga_5$ in $D$ dimensions would yield zero for this trace.
In order to calculate this type of diagrams, we have first evaluated the
triangle subgraph with the mathematically consistent definition of 
$\ga_5$ in $D$ dimensions according to \citeres{ga5HV,ga5BM} (here we
made use of the package {\sc Tracer}~\cite{tracer} for checking). 
After adding appropriate counterterms, which are necessary  to restore the
Ward identities, the result differs from the result obtained using a 
naively anticommuting $\ga_5$ only in terms proportional to the totally
antisymmetric tensor $\epsilon^{\mu\nu\rho\sigma}$. Inserting the
latter contribution into the two-loop diagrams, we find that the second
loop gives a finite contribution, so that it can be evaluated in four
dimensions without further complications.%
\footnote{For recent discussions of practical ways of treating
$\ga_5$ in higher-order calculations, see also \citeres{Jeg,impract}.}
The fermion line appearing in the 
second loop also yields an $\epsilon$-tensor contribution, which results,
after contraction with the $\epsilon$-tensor from the triangle subgraph,
in a non-vanishing contribution to the result for $\De r$.

As mentioned above, we perform the renormalization within the on-shell
scheme. It involves a one-loop subrenormalization of the Faddeev-Popov 
ghost sector of the theory, which is associated with the gauge-fixing part.  
The gauge-fixing part is kept invariant under renormalization.
For technical convenience, we manage this by a  
renormalization of the gauge parameters in such a way that it 
precisely cancels the renormalization of the parameters and fields in the 
gauge-fixing Lagrangian.%
\footnote{An alternative way of achieving that the gauge-fixing sector
does not give rise to counterterm contributions would have been to add the
gauge-fixing part to the Lagrangian only after renormalization, in which
case the renormalized gauge transformations would have to be used. }
To this end we have allowed
two different bare gauge parameters for both W and Z,
$\xi_1^{\PW,\PZ}$ and $\xi_2^{\PW,\PZ}$, and also mixing gauge parameters,
$\xiaz$ and $\xiza$. The renormalized parameters comply with the
$R_{\xi}$~gauge, with one free gauge parameter for each gauge boson.
With this prescription no counterterm contributions arise
from the gauge-fixing sector. Starting at the two-loop level,
counterterm contributions from the ghost sector have to be taken into
account in the calculation of physical amplitudes. They follow from the 
variation of the gauge-fixing terms $F^a$
under infinitesimal gauge transformations.
We have derived all the counterterms arising from the ghost
sector (extending the results of \citere{sbaugw1} to a
general $R_{\xi}$~gauge) and implemented them into the program \fea.
In this way we could verify the finiteness of individual 
(gauge-parameter-dependent) building blocks (e.g.\ the W- and the Z-boson 
self-energy) as a further check of the calculation.

Concerning the mass renormalization of unstable particles, from two-loop
order on it makes a difference whether the mass is defined according to
the real part of the complex pole of the S~matrix,
\beq
{\cal M}^2 = \overline{M}^2 - i \overline{M} \, \overline{\Gamma},
\label{eq:complpole}
\eeq
or according to the pole of the real part of the propagator. 
In \refeq{eq:complpole} ${\cal M}$ denotes 
the complex pole of the S~matrix and $\overline{M}$, $\overline{\Gamma}$ the 
corresponding mass and width of the unstable particle. We use the symbol
$\widetilde{M}$ for the real pole.

In the context of the present calculation, these 
considerations are relevant to the
renormalization of the gauge-boson masses, $\MW$ and $\MZ$. The two-loop
mass counterterms according to the definition of the mass as the real 
part of the complex pole are given by
\beqar
\de \overline{M}^2_{\PW, (2)} &=& 
     \mathrm{Re} \left\{\Sigma_{\rT, (2)}^{\PW}(\MW^2)\right\}
     - \de M^2_{\PW, (1)} \, \de Z_{(1)}^{\PW} + 
     \mathrm{Im}\left\{\Sigma_{\rT, (1)}^{\PW\prime}(\MW^2)\right\}
     \mathrm{Im}\left\{\Sigma_{\rT, (1)}^{\PW}(\MW^2)\right\} , \;
\label{eq:demw} \\
\de \overline{M}^2_{\PZ, (2)} &=& 
     \mathrm{Re} \left\{\Sigma_{\rT, (2)}^{\PZ\PZ}(\MZ^2)\right\}
     - \de M^2_{\PZ, (1)} \, \de Z_{(1)}^{\PZ\PZ} + 
     \frac{\MZ^2}{4} \left(\de Z_{(1)}^{\ga\PZ}\right)^2 + 
     \frac{\left(\mathrm{Im}\left\{
        \Sigma_{\rT, (1)}^{\ga\PZ}(\MZ^2)\right\}\right)^2}{\MZ^2} \non \\
  && {} + \mathrm{Im}\left\{\Sigma_{\rT, (1)}^{\PZ\PZ\prime}(\MZ^2)\right\}
     \mathrm{Im}\left\{\Sigma_{\rT, (1)}^{\PZ\PZ}(\MZ^2)\right\}
     \label{eq:demz}, 
\eeqar
where $\Sigma_{\rT, (1)}$, $\Sigma_{\rT, (2)}$ denote the transverse parts 
of the one-loop and two-loop self-energies (the terms from subloop 
renormalization are understood to be contained in the two-loop self-energies), 
and $\Sigma_{\rT, (1)}^{\prime}$ means the derivative of the one-loop
self-energy with respect to the external momentum squared. 
Field renormalization constants are indicated as $\de Z^V$.
The relations to the mass counterterms according to the real-pole definition,
$\de \widetilde{M}^2_{\PW, (2)}$ and $\de \widetilde{M}^2_{\PZ, (2)}$,
are given by
\beqar
\label{eq:demwdiff}
\de \overline{M}^2_{\PW, (2)} &=& 
   \de \widetilde{M}^2_{\PW, (2)} +
     \mathrm{Im}\left\{\Sigma_{\rT, (1)}^{\PW\prime}(\MW^2)\right\}
     \mathrm{Im}\left\{\Sigma_{\rT, (1)}^{\PW}(\MW^2)\right\},  \\
\de \overline{M}^2_{\PZ, (2)} &=& 
   \de \widetilde{M}^2_{\PZ, (2)} +
     \mathrm{Im}\left\{\Sigma_{\rT, (1)}^{\PZ\PZ\prime}(\MZ^2)\right\}
     \mathrm{Im}\left\{\Sigma_{\rT, (1)}^{\PZ\PZ}(\MZ^2)\right\} .
\label{eq:demzdiff}
\eeqar
It can easily be checked by direct computation that the terms in
\refeqs{eq:demwdiff}, (\ref{eq:demzdiff}) by which
the two definitions differ are gauge-parameter-dependent. Thus it is
obvious that at least one of the two prescriptions leads to a
gauge-dependent mass definition. While the problem of a proper
definition of unstable particles in gauge theories has already been addressed 
many times in the literature~\cite{unstab}, it should be noted that
in the present calculation two-loop contributions of the type
leading to a non-zero (and gauge-parameter-dependent) difference between
the two kinds of mass renormalization methods are for the first time fully 
included in a computation of a physical observable in the Standard Model. 
Explicitly, these are contributions from light fermions and bosonic loops 
evaluated in a general $R_{\xi}$~gauge. 
In the previous results for $\MW$, incorporating 
terms up to ${\cal O}(\GF^2 \Mt^2 \MZ^2)$~\cite{ewmt2} and $\MH$-dependent 
fermionic terms~\cite{ewmhdep}, the contribution 
$\mathrm{Im}\left\{\Sigma_{\rT, (1)}^{\prime}(M^2)\right\}
\mathrm{Im}\left\{\Sigma_{\rT, (1)}(M^2)\right\}$ was zero, 
making thus a strict distinction between the two mass definitions
unnecessary at the considered order.

Since our result has been obtained within a general $R_{\xi}$~gauge,
we can investigate the issue of whether the mass renormalization is
gauge-parameter-independent by explicit computation. In particular,
the two-loop counterterm to the weak mixing angle, $\de \swtwo$, ought
to be gauge-parameter-independent since $\sw$ is a physical observable
(note, however, that the same argument does not hold for the mass
counterterms of \refeq{eq:demw} and \refeq{eq:demz}; see e.g.\
\citere{bfm} for a discussion). We find that $\de \swtwo$ is only
gauge-parameter-independent with the definition of the gauge-boson
masses according to the complex pole, while the real-pole definition for
the masses leads to a gauge-parameter-dependent result for $\de \swtwo$.
This result is in accordance with what is expected from S-matrix theory, 
in which the complex pole 
is a gauge-invariant quantity~\cite{unstab}.


We have thus adopted the complex-pole definition as
given in \refeq{eq:demw} and \refeq{eq:demz}. Using this mass definition
leads to a Breit--Wigner parametrization of the resonance line shape with a
constant decay width. Experimentally the gauge-boson masses are
determined using a Breit--Wigner function with a running
(energy-dependent) width. Connecting the latter prescription with the
theoretical prediction involves the approximation
$\mathrm{Im}\left\{\Sigma^{\PW, \PZ}_{\rT, (1)}(s)\right\} \approx s
\Gamma_{\PW, \PZ}/M_{\PW, \PZ}$,
which is valid for the fermionic contributions to the W- and Z-boson
self-energies at one-loop order. As usual, $\Gamma_{\PW, \PZ}$ denote the W-
and Z-boson widths. As a consequence of the different Breit--Wigner
parametrizations, there is a numerical difference between the 
mass parameters corresponding to the definition used in the experimental
determination (denoted as $\MW$, $\MZ$ henceforth) and the mass
parameters in our calculation, $\overline{M}_{\PW}$,
$\overline{M}_{\PZ}$. The shift between these parameters is given
by~\cite{massshift}
$M_{\PW, \PZ} = \overline{M}_{\PW, \PZ} + \Gamma_{\PW, \PZ}^2/(2
M_{\PW, \PZ})$. 
Since $\MW$
and $\MZ$ enter on a different footing in our computation --- $\MZ$ is an 
experimental input parameter, while $\MW$ is calculated --- in order to 
evaluate the mass shifts we use the experimental value for the Z-boson width,
$\Gamma_{\PZ} = 2.944 \pm 0.0024$~GeV~\cite{mori00}, and the theoretical 
value for the W-boson width, which is given by
$\Gamma_{\PW} = 3 \GF \MW^3/(2 \sqrt{2} \pi) (1 +
2 \alps/(3 \pi))$ in sufficiently good approximation. This results in 
$\MZ \approx \overline{M}_{\PZ} + 34.1$~MeV and in the mass shifts
$\MW \approx \overline{M}_{\PW} + 27.4$~MeV and 
$\MW \approx \overline{M}_{\PW} + 27.0$~MeV for $\MW = 80.4$~GeV and 
$\MW = 80.2$~GeV, respectively.%
\footnote{
The difference in $\Gamma_{\PW}$ according to the way it is calculated,
through the tree-level result parametrized with $\al$, or the improved 
Born result parametrized with $\GF$, or the improved Born result including 
QCD corrections (which is the one we used), is formally of higher order 
(i.e.\ beyond ${\cal O}(\al^2)$) in the calculation of $\MW$. Its numerical 
effect is nevertheless not completely negligible; it changes the shift in 
$\MW$ by about $-2.9$~MeV if the tree-level result for
$\Gamma_{\PW}$ parametrized with $\al$ is used and by about $-1.4$~MeV
if the $\GF$~parametrization of the Born width (without QCD
corrections) is employed.}

\bigskip
We now turn to the numerical discussion of our result for $\De r$.
It should be noted that our definition of $\De r$ according to 
\refeq{eq:delr} is based on the expanded form $\left(1 + \De r\right)$ with 
$\De r = \De r^{(\al)} + \De r^{(\al^2)} + \ldots$ rather than on the resummed
form $1/(1 - \De r)$, indicating a resummation of leading one-loop 
contributions. The terms consistently taken into account 
at two-loop order with such a resummation
are explicitly contained in our two-loop contribution to $\De r$.
The result for $\De r$ contains the following contributions
\beq
\De r = \De r^{(\al)} + \De r^{(\al\alps)} + \De r^{(\al\alps^2)} +
\De r^{(N_{\mathrm{f}} \al^2)} + \De r^{(N_{\mathrm{f}}^2 \al^2)}, 
\label{eq:delrcontribs}
\eeq
where $\De r^{(\al)}$ is the one-loop result, \refeq{eq:delrol}, $\De
r^{(\al\alps)}$ and $\De r^{(\al\alps^2)}$ are the two-loop~\cite{qcd2}
and three-loop~\cite{qcd3} QCD corrections, while 
$\De r^{(N_{\mathrm{f}} \al^2)}$ is the new electroweak two-loop result. 
The notation $(N_{\mathrm{f}} \al^2)$ symbolizes the contribution of all 
diagrams containing one fermion loop, where $N_{\mathrm{f}}$ stands both for 
the top/bottom contribution and for all light-fermion species. The term 
$\De r^{(N_{\mathrm{f}}^2 \al^2)}$ contains the pure fermion-loop
contributions in two-loop order. 
Since the pure fermion-loop contributions in three- and four-loop order
have been found to be numerically small, as a consequence of accidental
numerical cancellations, with a net effect of only about 1~MeV in $\MW$
(using the real-pole definition of the gauge-boson masses)~\cite{floops},
we have not included them here.

\begin{figure}[ht]
\vspace{1em}
\begin{center}
\epsfig{figure=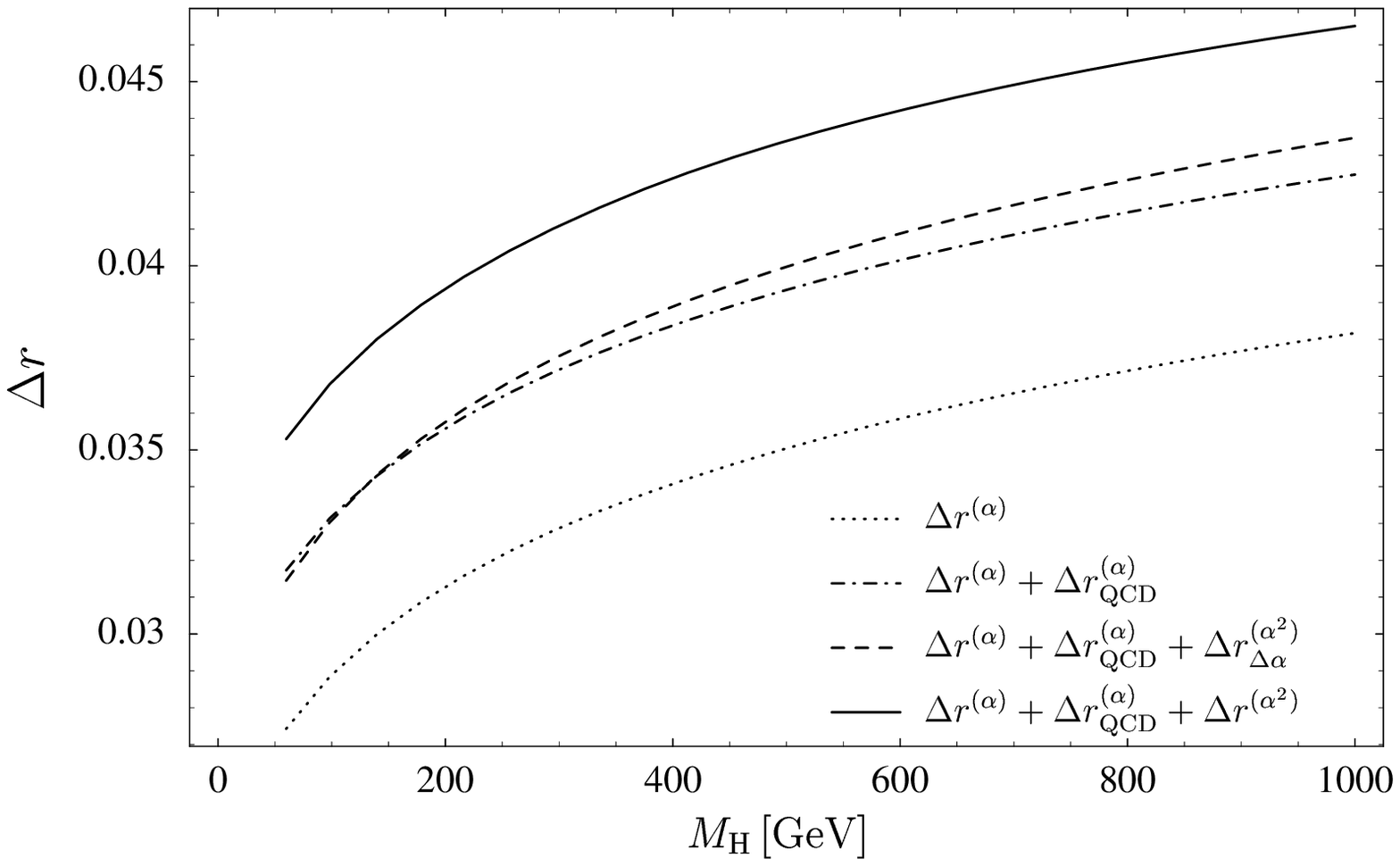,width=14cm}
\end{center}
\caption[]{{\small
Different contributions to $\De r$ as a function of $\MH$.
The one-loop contribution, $\De r^{(\al)}$, is supplemented by the
two-loop and three-loop QCD corrections, 
$\De r^{(\al)}_{\mathrm{QCD}} \equiv \De r^{(\al\alps)} + \De
r^{(\al\alps^2)}$,
and the fermionic electroweak two-loop contributions,
$\De r^{(\al^2)} \equiv \De r^{(N_{\mathrm{f}} \al^2)} +
\De r^{(N_{\mathrm{f}}^2 \al^2)}$.
For comparison, the effect of the two-loop corrections induced by a
resummation of $\De\al$, $\De r^{(\al^2)}_{\De\al}$, is shown
separately.
}}
\label{fig:delrcontr}
\end{figure}

In \reffi{fig:delrcontr} the different contributions to $\De r$ are shown 
as a function of $\MH$. Here $\MW$ is kept fixed at its experimental 
central value, $\MW = 80.419$~GeV, and $\mt = 174.3$~GeV~\cite{pdg}
is used. 
The effects of the QCD corrections, of the two-loop corrections induced by
a resummation of $\De \al$, and of the purely electroweak fermionic 
two-loop corrections are shown separately. The purely electroweak 
two-loop contributions are sizeable and amount to about 10\% of the
one-loop result.
We have compared the Higgs-mass
dependence of $\De r$ with the result previously obtained in 
\citere{ewmhdep} and found perfect agreement.

\begin{figure}[ht]
\vspace{1em}
\begin{center}
\epsfig{figure=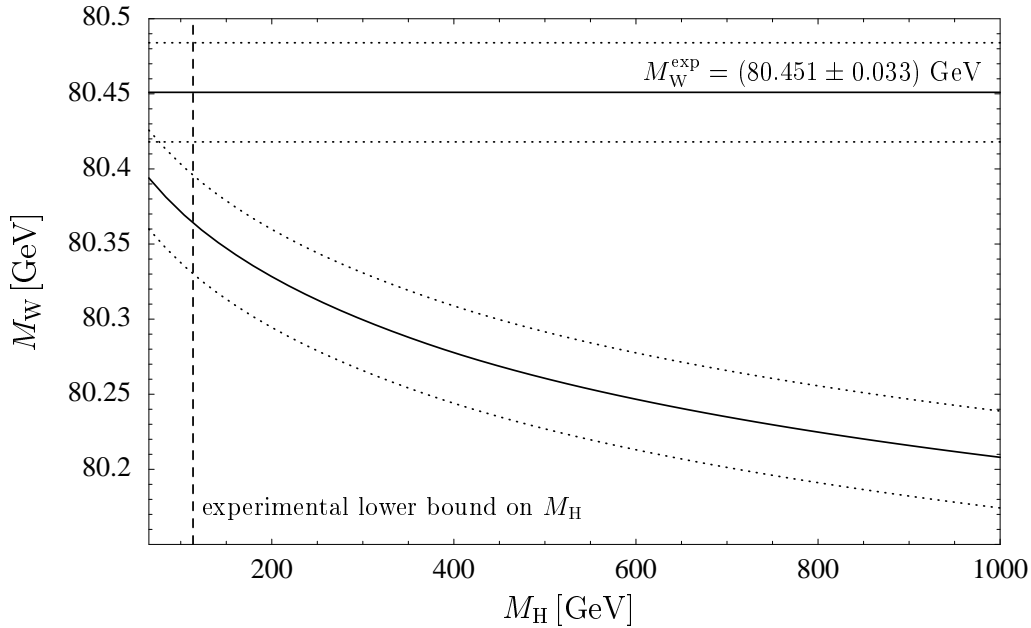,width=14cm}
\end{center}
\caption[]{{\small
The SM prediction for $\MW$ as a function of $\MH$ for 
$\mt = 174.3 \pm 5.1$~GeV is compared with the current experimental
value, $\MW^{\mathrm{exp}} = 80.419 \pm 0.038$~GeV~\cite{mori00}, and 
the experimental 95\% C.L.\ lower bound on the Higgs-boson mass,
$\MH = 107.9$~GeV~\cite{mhlimit}.
}}
\label{fig:MWpred}
\end{figure}

The prediction for $\MW$ is obtained from the input parameters by solving 
\refeq{eq:delr}. Since $\De r$ itself depends on $\MW$ this is
technically done using an iterative procedure.
The prediction for $\MW$ based on the results of \refeq{eq:delrcontribs}
is shown in \reffi{fig:MWpred} as a function of $\MH$ for $\mt = 174.3
\pm 5.1$~GeV~\cite{pdg} and $\De\al = 0.05954 \pm
0.00065$~\cite{eidjeg}. The current experimental value,
$\MW^{\mathrm{exp}} = 80.419 \pm 0.038$~GeV~\cite{mori00}, and 
the experimental 95\% C.L.\ lower bound on $\MH$ 
($\MH = 107.9$~GeV~\cite{mhlimit}) from the direct search
are also indicated. The plot shows the well-known preference for a light
Higgs boson within the SM. Confronting the theoretical prediction
(allowing a variation of $\mt$, which at present dominates the
theoretical uncertainty, and $\De\al$ within $1\sigma$) with the 
$1\sigma$ region of $\MW^{\mathrm{exp}}$ and the 95\% C.L.\ lower bound 
on $\MH$, only a rather small region in the plot (corresponding to 
$107.9 \gev < \MH \lsim 140 \gev$) matches all three 
constraints.

We have compared our results with those of an expansion for asymptotically 
large values of $\mt$ up to 
${\cal O}(\GF^2 \Mt^2 \MZ^2)$~\cite{ewmt2,gambpriv}. 
The results are shown in \refta{tab:MWcomp1} for different values of
$\MH$. For the input parameters the values of \citere{ewmt2} have been
chosen, i.e.\ $\mt = 175$~GeV, $\MZ = 91.1863$~GeV, $\De\al = 0.0594$,
$\alps(\MZ) = 0.118$. Relatively good agreement is found, with maximal 
deviations in $\MW$ of about $5$~MeV. If we had chosen a different 
parametrization of $\Ga_{\PW}$ in the above calculation of the shift 
between the masses
corresponding to the fixed and the running width definition, a somewhat
larger deviation to the result of \citere{ewmt2} would have been obtained.

\btab
$$
\begin{array}{|c||c|c||c|} \hline
\MH /\GeV &
\MW/\GeV &
\MW^{\mathrm{expa}} /\GeV &
\De \MW /\MeV\\ \hline
65   & 80.3985 & 80.4039 & -5.4 \\ \hline
100  & 80.3759 & 80.3805 & -4.6 \\ \hline
300  & 80.3039 & 80.3061 & -2.2 \\ \hline
600  & 80.2509 & 80.2521 & -1.2 \\ \hline
1000 & 80.2122 & 80.2129 & -0.7 \\ \hline
\end{array}
$$
\caption{The two-loop result for $\MW$ based on \refeq{eq:delrcontribs}
is compared with the results of an expansion in $\mt$ up to ${\cal
O}(\GF^2 \Mt^2 \MZ^2)$~\cite{ewmt2,gambpriv}, $\MW^{\mathrm{expa}}$.
The last column indicates the difference between the two results.
\label{tab:MWcomp1}}
\etab

The deviations in the last column of \refta{tab:MWcomp1}
can of course not be attributed exclusively to differences in the 
two-loop top-quark and light-fermion
contributions, because the results also differ 
by a slightly different treatment of those higher-order terms that are
not yet under control, such as purely bosonic two-loop contributions and
effects from scheme dependence. 
A detailed discussion of those differences and of the remaining theoretical 
uncertainties from unknown higher-order corrections will be given in a 
forthcoming publication.


Following \citere{DGPS}, we also provide a simple numerical
parametrization of our result for $\MW$. It is given by
\beq
\MW = \MW^0 - c_1 \, \mathrm{dH} - c_5 \, \mathrm{dH}^2 + c_6 \, \mathrm{dH}^4 
       - c_2 \, \mathrm{d}\al + c_3 \, \mathrm{dt} 
       - c_7 \, \mathrm{dH} \, \mathrm{dt} - c_4 \, \mathrm{d}\alps ,
\label{eq:simppar}
\eeq
where 
\beq
\mathrm{dH} = \ln\left(\frac{\MH}{100 \gev}\right), \; 
\mathrm{dt} = \left(\frac{\mt}{174.3 \gev}\right)^2 - 1, \; 
\mathrm{d}\al = \frac{\De\al}{0.05924} - 1, \;
\mathrm{d}\alps = \frac{\alps(\MZ)}{0.119} - 1,
\eeq
and $\MZ = 91.1875$~GeV~\cite{mori00} has been used.
For the coefficients $c_1, \ldots, c_7$ we have obtained via a 
least squares fit
$\MW^0 = 80.3755$~GeV, $c_1 = 0.05613$, $c_2 = 1.081$, $c_3 = 0.5235$,
$c_4 = 0.0763$, $c_5 = 0.00936$, $c_6 = 0.000546$, $c_7 = 0.00573$.
The parametrization of \refeq{eq:simppar} approximates our full result
for $\MW$ within $0.4$~MeV for $65 \gev \leq \MH \leq 1 \tev$.

\bigskip
In summary, we have evaluated the complete fermionic two-loop
contributions to the W-boson mass within the electroweak Standard Model.
Our result improves on previous results as it does not involve any
approximations in the top-quark and the Higgs-boson mass and also
contains the contributions of all light fermions in the Standard Model.
Within our calculation we have defined the gauge-boson masses according
to the complex pole of the S~matrix, which ensures the gauge-parameter
independence of the mass definition.
We have provided a simple numerical parametrization of our result, which
approximates the full result with sufficient accuracy for all values of
$\MH$ up to 1~TeV. In comparison with the previous result obtained for 
$\MW$ by an expansion for asymptotically large values in $\mt$ up to
next-to-leading order we find slightly lower values of $\MW$, sharpening
thus the tendency towards a light Higgs boson within the Standard
Model.

\bigskip

\vspace{- .3 cm}
\section*{Acknowledgements}

We thank S.~Bauberger, M.~B\"ohm, S.~Heinemeyer and A.~Stremplat for
collaboration in early stages of this work. We also thank 
P.~Gambino and D.~St\"ockinger for useful discussions and communications.
We are grateful to M.~Awramik and M.~Czakon for detailed comparisons with their
results \cite{AC2lf}, which helped to identify an error in our computation.

\end{document}